\documentclass[reprint]{JASAnew}
\usepackage{amsfonts}
\usepackage{natbib}
\usepackage[margin=1in]{geometry}
\usepackage{url}
\usepackage{amsmath}
\usepackage{amsthm}
\usepackage{graphicx}
\usepackage{float}
\usepackage{booktabs}

\begin{document}


\title[Estimating radii of air bubbles in water]{Estimating the radii of air bubbles in water using passive acoustic monitoring}

\author{P. Hubert}
\affiliation{Technology and Data Science Department, EAESP - FGV \\ Acoustics and Environment Lab - EP - USP}
\author{L. Padovese}
\affiliation{Acoustics and Environment Lab - EP - USP}

\thanks{The authors gratefully acknowledge University of S\~{a}o Paulo and support from SHELL Brazil (subsidiary company of Royal Dutch Shell) and FAPESP, through the Research Centre for Gas Innovation (RCGI) hosted by the University of S\~{a}o Paulo (FAPESP Grant Proc. 2014/50279-4). We would also like to thank FAPESP and CNPq for their support, by grants number FAPESP 2016/02175-0  and CNPq 303992/2017-4.}


\begin{abstract}
The study of the acoustic emission of underwater gas bubbles is a subject of both theoretical and applied interest, since it finds an important application in the development of acoustic monitoring tools for detection and quantification of underwater gas leakages. An underlying physical model is essential in the study of such emissions, but is not enough: also some statistical procedure must be applied in order to deal with all uncertainties (including those caused by background noise). In this paper we take a probabilistic (Bayesian) methodology which is well known in the statistical signal analysis communitiy, and apply it to the problem of estimating the radii of air bubbles in water. We introduce the \emph{bubblegram}, a feature extraction technique graphically similar to the traditional spectrogram but tailored to respond only to pulse structures that correspond to a given physical model. We investigate the performance of the bubblegram and our model in general using laboratory generated data.

\end{abstract}

\maketitle

\section{Introduction}\label{sec:intro}

Underwater gas leakages are a matter of concern for several industries. Detecting and quantifying such leakages is thus a problem with theoretical interest but also with important engineering implications.

In this paper we are particularly interested in the use of Passive Acoustic Monitoring (PAM) technologies. This approach is cheaper compared to other, more complex options involving active monitoring, but presents the challenge of analyzing the acquired signal which will be contaminated by background noise and other acoustic events at a wide frequency range.

In this work we consider the single sensor case and propose a probabilistic (Bayesian) approach to the problem of modelling the acoustic behavior of underwater air bubbles and bubble plumes. Our approach is independent of physical models; we adopt the model of \citep{Strasberg1956} as our first choice in this paper, but generalization for other models is straightforward.

Based on this approach we propose a graphical evaluation tool, the \emph{bubblegram}, which bears resemblance to the usual spectrogram graphs widely used in signal processing. However, instead of representing the signal's energy at different frequency ranges as the spectrogram does, the bubblegram represents the posterior density for the presence of bubbles of any radius $R$ at a given time $t$. 

The paper is organized as follows: section \ref{sec:prob} describes the Bayesian methodology for the analysis of exponentially decaying sinusoids in noisy signals. Secion \ref{sec:bubbles} describe the physical model of \citep{Strasberg1956} and shows how to plug this model into the probabilistic framework. Section \ref{sec:inference} presents the bubblegram and section \ref{sec:results} presents the results of our model in laboratory data. Section \ref{sec:conc} concludes the paper.

\section{Probabilistic modelling}\label{sec:prob}

Consider a real-valued signal along a single time dimension, $y(t)$, sampled (not necessarily uniformly) at points $\{t_i\}$. We adopt the notation $y_t$ for the discretely sampled signal.

Define a pulse as an exponentially decaying sinusoidal function

\begin{equation}
x_t = A\cdot e^{-\lambda t}\cdot cos(\omega t  + \phi)
\end{equation}

where $A$ is the (assumed constant) amplitude, $\lambda$ is the decay factor, $\omega$ is the frequency and $\phi$ the pulse phase. 

This equation assumes that $t=0$ is the starting time of the pulse. 

Instead of writing the pulse equation depending non-linearly on the phase $\phi$, we adopt the parameterization

\begin{equation}\label{eq:pulse}
x_t = e^{-\lambda (t-t_0)}\left( A\cdot cos(\omega (t-t_0)) + B\cdot sin(\omega (t-t_0))\right)\cdot \mathbf{1}_{t > t_0}
\end{equation}

which is equivalent to the former but replaces the nonlinear parameter $\phi$ by a new, linear parameter (the amplitude $B$). Also we included the parameter $t_0$ to generalize the model for pulses starting at times different from $t=0$, and the function $\mathbf{1}_{U}$ is the indicator function, taking the value $1$ when $U$ is true, and $0$ otherwise.

The acquired signal $y_t$ is assumed to be formed by a single pulse starting at time $t_0$, corrupted with Gaussian white noise $r_t$, with $E\left(r_t\right) = 0$ and $var(r_t) = \sigma^2$.

\begin{equation}\label{eq:mod1}
y_t = x_t  + r_t
\end{equation}

In this general form the model can be used to describe many different phenomena. Its main interest lies in the fact that it explicitly models the pulse in terms of decay constant and fundamental frequency, as opposed to traditional spectral analysis that focus on the frequency alone. This model, being probabilistic in nature, naturally incorporates uncertainty into the analysis, and also allows the inclusion of any prior information available about the signal and the noise.

In this work we are particularly interested in the situation where the decay constant $\lambda$ and the fundamental frequency $\omega$ are related by a known, deterministic function. This is the case when the pulse is taken to represent the acoustic emission of a gas bubble in water, in which case both $\lambda$ and $\omega$ depend on the bubble's mean (equilibrium) radius. Directly estimating this radius can aid the design of signal detectors for underwater gas leakages, and also serves as a tool for estimation of the leaked gas flow.

Now taking the structural model of equation \ref{eq:mod1}, and assuming Gaussian white noise, the log-likelihood of the model can be written as

\begin{equation}\label{eq:lhood}
\ell(y_t \mid A, B, \omega, \sigma, \lambda, t_0) = -\frac{N}{2}log\left(2\pi \sigma^2\right)-\frac{1}{2\sigma^2}\sum_{i=1}^N \left(y_t - x_t\right)^2
\end{equation}

Since our main interest lies in the estimation of $\lambda$, $\omega$ and $t_0$, we proceed to marginalize all other factors from the likelihood. By adopting uniform priors for the amplitudes $A$ and $B$, and a Jeffreys' prior for $\sigma^2$, this marginalization can be performed analitically, resulting in the following expression for the marginalized likelihood:

\begin{align}\label{eq:mlhood}
\begin{split}
P(y_t \mid t_0, \omega, \lambda) \propto & \left(\sum_{i=1}^Nd_i\right)^{-1/2}\left(\sum_{i=1}^Ne_i\right)^{-1/2} \times \\ 
& \left(\sum_{i=1}^Ny_i^2 - \frac{\left(\sum_{i=t_0}^N y_id_i\right)^2}{\sum_{i=t_0}^Nd_i^2} - \frac{\left(\sum_{i=t_0}^N y_ie_i\right)^2}{\sum_{i=t_0}^Ne_i^2} \right)
\end{split}
\end{align}

with

\begin{align}
\begin{split}
& d_i = e^{-\lambda(t-t_0)}cos(\omega (t_i - t_0)) \\
& e_i = e^{-\lambda(t-t_0)}sin(\omega (t_i - t_0))
\end{split}
\end{align}

This expression can further on be written in a concise manner as

\begin{equation}\label{eq:post}
P(y_t \mid t_0, \omega, \lambda) \propto K(\omega, \lambda, t_0)\left[\lVert y \rVert^2 - \left(\frac{\langle y, d \rangle}{\lVert d \rVert}\right)^2  - \left(\frac{\langle y, e \rangle}{\lVert e \rVert}\right)^2\right]
\end{equation}

Notice that this expression is independent of the form of both $d_i$ and $e_i$, given that these functions do not depend on the marginalized parameters. 

This methodology is not new, as it has been originally proposed and studied by \citep{Bretthorst1988, Bretthorst1990A, Bretthorst1990B, Bretthorst1990C} and applied to the analysis of Nuclear Magnetic Ressonance data. Here we apply  it to the analysis of the acoustic emission of air bubbles in water.

\section{Acoustic emission of air bubbles}\label{sec:bubbles}

The study of the acoustic emission of air bubbles in liquid media dates back to \citep{Minnaert1933}, is treated again by \citep{Strasberg1956}, and more recently has been extensively studied by \citep{Leighton1996, Ainslie2011, Leighton2012B} and others. 

In \citep{Strasberg1956} a formula relating the bubble's fundamental frequency of oscilation to its radius is given as:


\begin{equation}\label{eq:freq}
\omega_0 = \sqrt{\frac{3\gamma P_0}{\rho_0}}\left(\frac{1}{2\pi R_0}\right)
\end{equation}

here $R_0$ is the bubble steady-state radius; $\rho_0$ is the liquid's density; $\gamma$ is the ratio of specific heats, and $P_0$ the static pressure. 


The model for the pressure variations on the surface of a single bubble over time is of the same form as \ref{eq:pulse}, given by:

\begin{equation}\label{eq:leighton}
x_t = \left(\omega_0R_0\right)^2\frac{\rho_0}{r}\alpha_0e^{-\omega_0\lambda(t_0-t)}\mathbf{1}_{t > t_0}cos\left(\omega_0(t_0-t)\right)
\end{equation}

with $\alpha_0$ the initial wall amplitude of the bubble, and $r$ the distance between the bubble and the sensor. 


By assuming constant the liquid's density, the distance between the bubble and the sensor, and the initial amplitude, we rewrite equation \ref{eq:leighton} as

\begin{equation}\label{eq:newpulse}
x_t = C\cdot\left(\omega_0R_0\right)^2e^{-\omega_0\lambda(t-t_0)}\mathbf{1}_{t > t_0}cos\left(\omega_0(t-t_0)\right)
\end{equation}

Now including a phase parameter and reparameterizing, we see that model \ref{eq:newpulse} can be put in the form of \ref{eq:mlhood} with

\begin{align}
\begin{split}
& d_i = \left(w_0R_0\right)^2e^{-\omega_0\lambda(t-t_0)}cos\left(\omega_0(t-t_0)\right) \\
& e_i = \left(w_0R_0\right)^2e^{-\omega_0\lambda(t-t_0)}sin\left(\omega_0(t-t_0)\right)
\end{split}
\end{align}

Using equation \ref{eq:freq} that gives $\omega_0$ as a function of $R_0$, $\omega_0 = \Omega(R_0)$, we can eliminate $\omega$ from the model:

\begin{equation}\label{eq:postfinal}
P(y_t \mid t_0, R_0, \lambda) \propto K(t_0, R_0, \lambda)\left[\lVert y \rVert^2 - \left(\frac{\langle y, d \rangle}{\lVert d \rVert}\right)^2  - \left(\frac{\langle y, e \rangle}{\lVert e \rVert}\right)^2\right]
\end{equation}

with

\begin{align}
\begin{split}
& d_i = \left(\Omega(R_0)R_0\right)^2e^{-\Omega(R_0)\lambda(t-t_0)}cos\left(\Omega(R_0)(t-t_0)\right) \\
& e_i = \left(\Omega(R_0)R_0\right)^2e^{-\Omega(R_0)\lambda(t-t_0)}sin\left(\Omega(R_0)(t-t_0)\right)
\end{split}
\end{align}

Finally, \citep{Strasberg1956} also provides an expression for the decay constant as a function of the fundamental frequency. In our notation

\begin{equation}\label{eq:decay}
\lambda = 0.014 + 1.1\cdot 10^{-5}\Omega(R_0)
\end{equation}

By plugging this equation into our model we are able to eliminate one more parameter, writing $\lambda = \Lambda(R_0)$, and arriving at a final model depending only on the bubble's radius $R_0$ and its time of occurrence, $t_0$.

This elimination depends crucially on the physical models relating the frequency $\omega$ and decay factor $\lambda$ to the bubble's radius. In this formulation we accepted these models as groundtruth, and did not include any uncertainty in these functions. It is possible, however, to augmentate the model by including error terms in the equations for $\omega$ and $\lambda$, and also to include prior distributions for all physical constants involved in the model. For this work, however, we chose to adopt the simplest formulation to allow a direct comparison of our model with the traditional tools of signal analysis.

\section{Estimating the bubble radius}\label{sec:inference}

An immediate use for this model is the estimation of $R_0$ and $t_0$ from a given signal. This can be performed by first adopting prior distributions for each parameter in the model, and obtaining a joint posterior by application of Bayes' theorem. 

To illustrate this process we generate random data for a single pulse representing a bubble starting at $t_0 = 0.5$ with radius $R_0 = 1mm$. We add Gaussian white noise with $\sigma^2 = 0.1$. 

\begin{figure*}
\figline{\fig{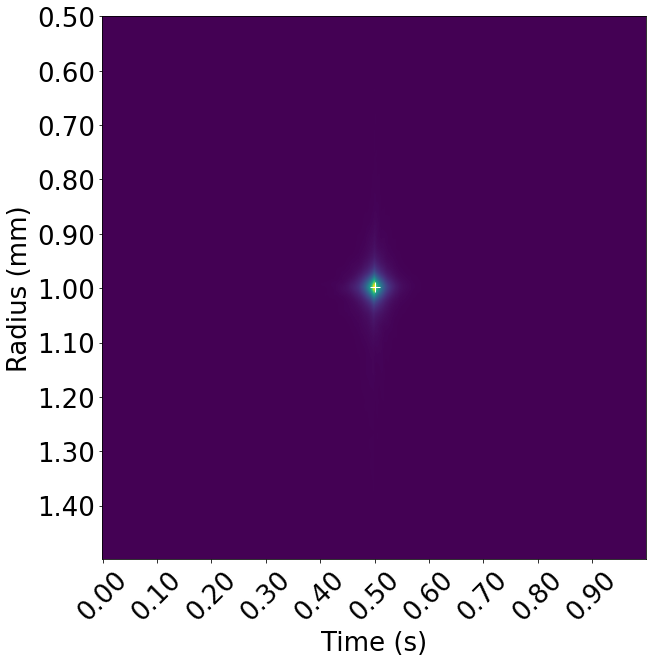}{.4\textwidth}{(a) Log-posterior}
\fig{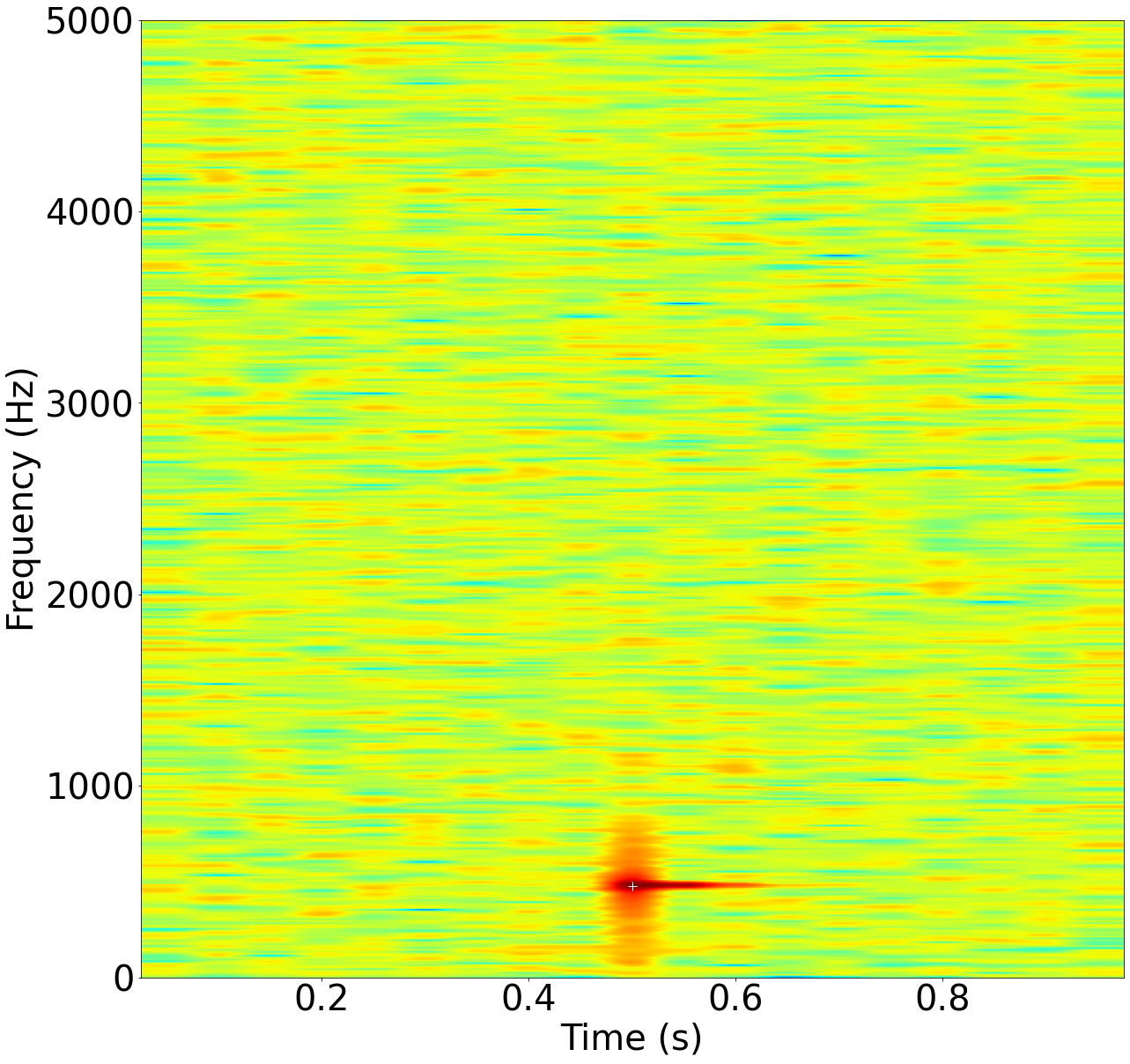}{.4\textwidth}{(b) Spectrogram}}
\caption{\label{fig:fig1} Log-posterior and spectrogram of simulated, single pulse data.}
\end{figure*}

By assuming flat, uninformative priors for all parameters, we obtain that the (log) joint posterior is equal (up to a constant) to the (marginal) log-likelihood from equation \ref{eq:postfinal}. The values of this log-posterior over a grid of $500\times 500$ points in the $time$ $\times$ $radius$ plane appear as the color scale in figure \ref{fig:fig1}, where brighter colors are associated with high posterior mass for a bubble with a given radius at a given time. The true values of $t_0$ and $R_0$ are marked with a white cross.

In figure \ref{fig:fig1} we also included a spectrogram of the same data for comparison. 

The first and most important difference between the visualization of the log-posterior and the spectrogram is the noise. Since the log-posterior model incorporates the noise term explicitly in the calculations (via the marginalization of the noise power $\sigma^2$) it is able to better recover the signal parameters. The spectrogram in itself is not a statistical procedure, and thus does not account for the presence of noise. The consequence is that signal and noise get blurred, as we see in the figure.

It will be also interesting to analyze a signal containing a sequence of bubbles of various sizes. This will occurr in most practical situations with higher gas flows. The original model we proposed in section \ref{sec:prob} assumes a single bubble; however, the presence of the parameter $t_0$ allows the application of this same model for a longer signal. One can think of the usual Short-Term Fourier Transform approach, where a local model is applied to small sections of a long signal, possibly after applying a smoothing function in the time domain. The smoothing, in our model, would be represented by a prior distribution on $t_0$.

What we show in the next figures is the plot of the posterior log-density for $R_0$, conditional on a fixed value for $t_0$. In this sense we are looking for the radii with larger evidence of appearing in a pulse starting at time $t$, for $t \in \left[t_a,t_b\right]$. 

When the gas flow is large enough and the bubbles start to cluster together, the hydrodinamical models for the acoustic emission of a single bubble are no longer exactly valid \citep{Ainslie2011}. However, since the acoustic emission of a gas bubble in water depends only on the forming process of the bubble, spurious acoustic emissions will occur only in the events of fragmentation of a bubble or the coalescence of two or more bubbles. For these reasons we believe that our model can be applied succesfully even to the analysis of higher flow bubble plumes.

\begin{figure*}
\figline{\fig{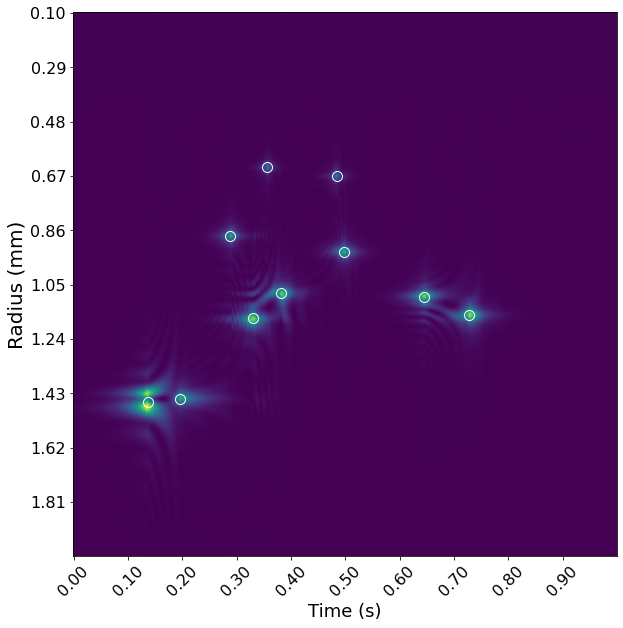}{.4\textwidth}{(a) Log-posterior}
\fig{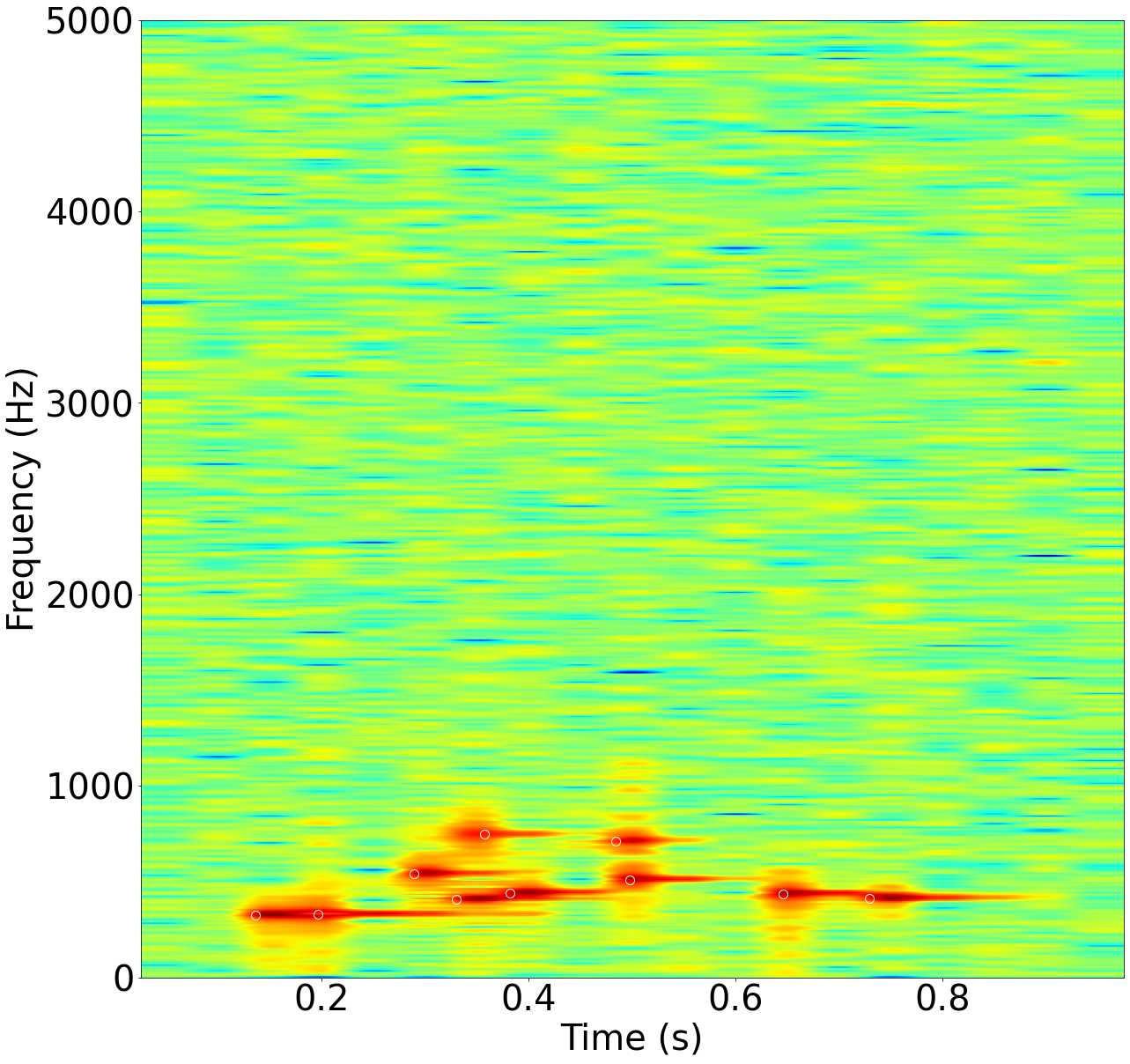}{.4\textwidth}{(b) Spectrogram}}
\caption{\label{fig:fig2} Log-posterior and spectrogram for simulated data.}
\end{figure*}

To understand the behavior of the log-posterior in this case we simulate a sequence of $10$ pulses occurring along a one second interval. The bubbles' starting times and radii are simulated from uniform distributions between $[0,1]$ and $[0.5, 1.5]$, respectively. The evaluation of the posterior over a grid of $500 \times 500$ points and $R_0 \in [0.2, 2]$ and $t_0 \in [0, 1]$ appear in figure \ref{fig:fig2}, where we included empty circles centered at the true values of each bubble. Again the spectrogram estimated from the same signal is included for comparison.

Both figures \ref{fig:fig1} and \ref{fig:fig2} show that the model behaves as expected. In the multiple bubbles case it appears that larger bubbles might be easier to detect (since the log-posterior is higher around these bubbles), but local peaks in the log-likelihood can be found near the radius and time of occurrence of all pulses. By comparing the log-posterior plots with the corresponding spectrograms we see that the log-posterior not only directly estimates the bubble's radii but also is more precise, since the model explicitly incorporates the background noise as already observed.

These plots that evaluate the joint posterior of $R_0$ and $t_0$ over a (not necessarily uniform) grid show some visual resemblance with the traditional spectrogram plots usually applied to the analysis of acoustic signals. Inspired by this similarity we call the plots of the log-posterior  \emph{bubblegrams}, since they are designed to respond to the acoustic emission of air bubbles in water, representing the posterior probability associated with a given radius at a given time. One advantage of the bubblegram over the traditional spectrogram is that it does not suffer from the usual resolution limitations of the spectrogram; to understand how that can be  the case, one can imagine that by the use of the probabilistic model one is effectively introducing an \emph{infinite zero-padding} into the signal, thus allowing any arbitrary level of resolution both in time and radius (frequency) domains.

\section{Experimental results}\label{sec:results}

In this section we apply our model to the analysis of experimental data. The experiments were conducted by the \emph{Acoustics and Environment Laboratory} (LACMAM in the portuguese acronym) at the Polytechnic School, University of S\~{a}o Paulo.

\begin{figure*}
\figline{\fig{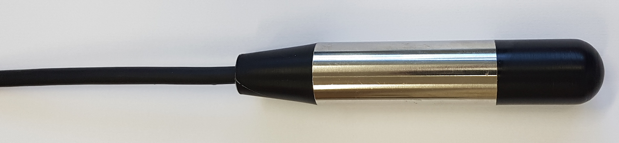}{.5\textwidth}{}}
\caption{\label{fig:fig3} Hidrophone developed by LACMAM.}
\end{figure*}

We used sensors (hydrophones) developed by the laboratory itself; these sensors have a frequency band of $5$ $Hz$ to $60$ $kHz$, with sensitivity of $-157 \pm 2$ $dB$ $rel$ $1\mu Pa$ (preamplified) or $-203 \pm 2$ $dB$ $rel 1 \mu Pa$ (not preamplified). A picture of one sensor appears in figure \ref{fig:fig3}.

\begin{figure*}
\figline{\fig{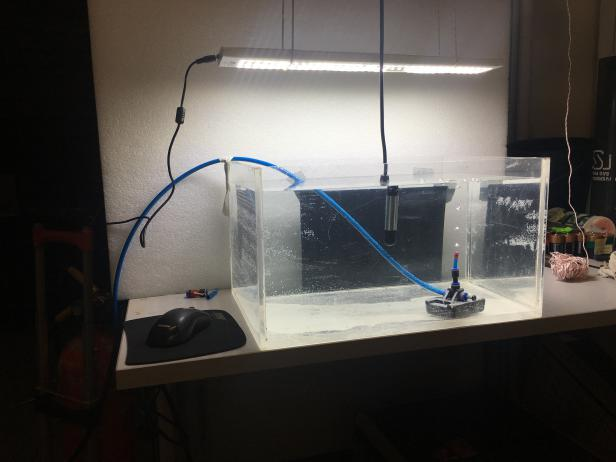}{.5\textwidth}{}}
\caption{\label{fig:fig4} Setup for measuring bubble sizes.}
\end{figure*}

Two experiments were conducted. The first was designed to measure the size of single bubbles by using a high speed CCD camera with a $25$ $mm$ lens with $F1.4$ opening. For this experiment a small aquarium with dimensions $45$ $cm$ (length) $\times$ $40$ $cm$ (width) $\times$ $25$ $cm$ (height) was used. Three nozzles with diameters $2.5$, $4$ and $6$ $mm$ were positioned one at a time under a $22$ $cm$ water column, and a flow of less than $1$ $l/min$ of air was induced by the use of a flow controller connected to a compressed air cylinder. A picture of the setup is seen on figure \ref{fig:fig4}.

\begin{figure*}
\figline{\fig{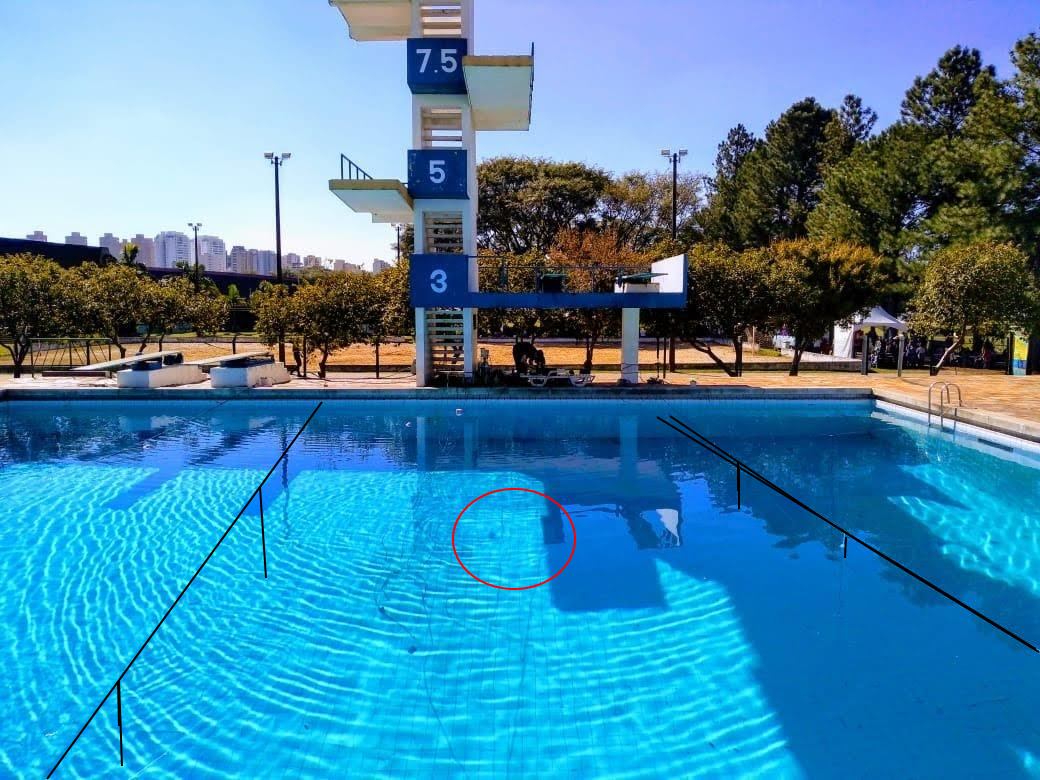}{.5\textwidth}{}}
\caption{\label{fig:fig5} Diving tank; see text for details.}
\end{figure*}

\begin{figure*}
\figline{\fig{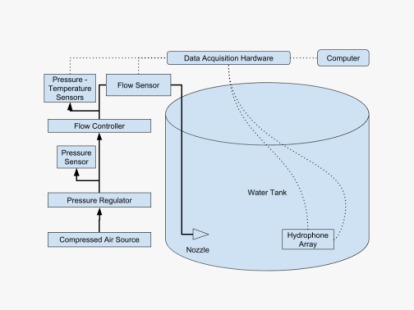}{.5\textwidth}{}}
 \caption{\label{fig:fig6} Schematic drawing of the experimental setup.}
\end{figure*}

The second experiment was designed to obtain acoustic signals emitted by bubble plumes in water. It was performed in a diving tank with a depth of $5$ $m$. The bubbles are created by the injection of compressed air into the tank, with flow, pressure and exit diameter orifice controlled. Figures \ref{fig:fig5} and \ref{fig:fig6} show the equipment used in this experiment. In figure \ref{fig:fig5} the black lines show the position of the four sensors used, and the red circle indicates the approximate position of the flow controller nozzle.

%

%

\subsection{Data acquisition and experimental results}

\subsubsection{Single bubble measurements}

We recorded high speed ($1000$ frames per second) videos of the bubbles generated in the aquarium, with a ruler attached to the aquarium wall. Figure \ref{fig:fig7} below shows snapshots of bubbles generated with the three different nozzle diameters. The labels in each picture indicate the \textbf{diameter of the nozzle} used in the air injection.

\begin{figure}[!ht]
\figline{\fig{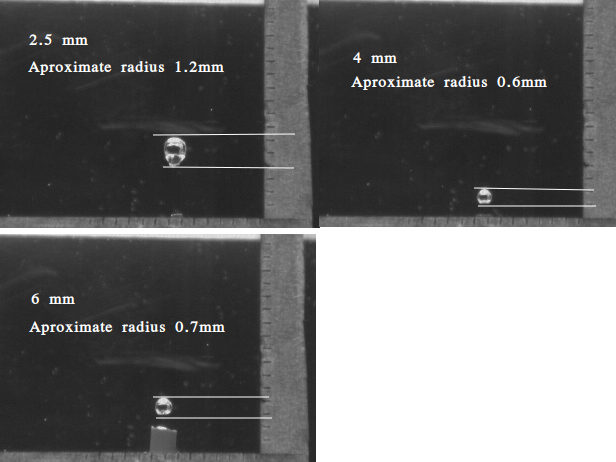}{.5\textwidth}{}}
 \caption{Images of single bubbles}
 \label{fig:fig7}
\end{figure}

The acoustic emissions were recorded with a sampling frequency of $48$ $kHz$. To create the bubblegrams we first passed the signals through a Butterworth bandpass filter (with pass range $200$ to $3000$ $Hz$). We then selected the sections of the signal corresponding to the photographed bubbles. These preprocessing steps were taken in order to accelerate further processing of the signal.

\begin{figure}[H]
\figline{\fig{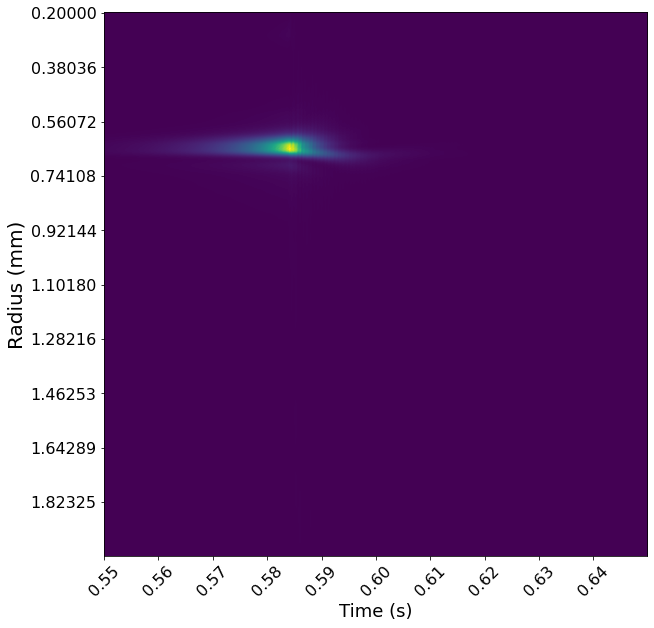}{.4\textwidth}{(a)$2.5$ $mm$ nozzle}
\fig{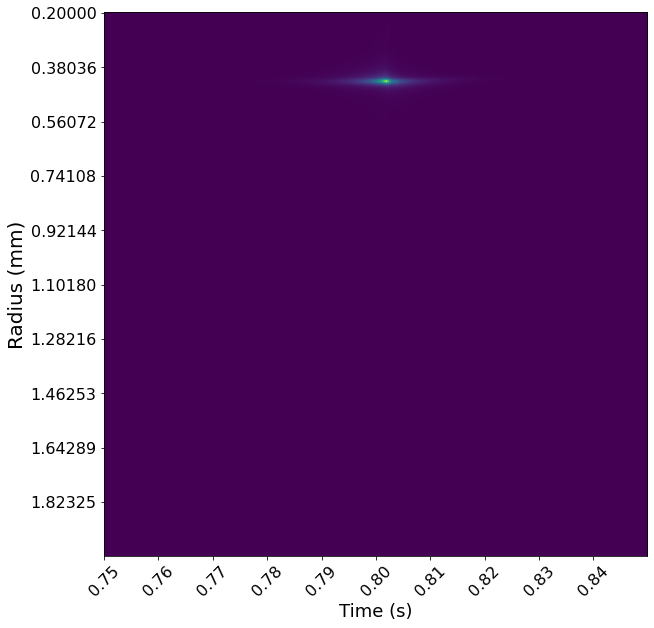}{.4\textwidth}{(b) $4$ $mm$ nozzle}}
\figline{\fig{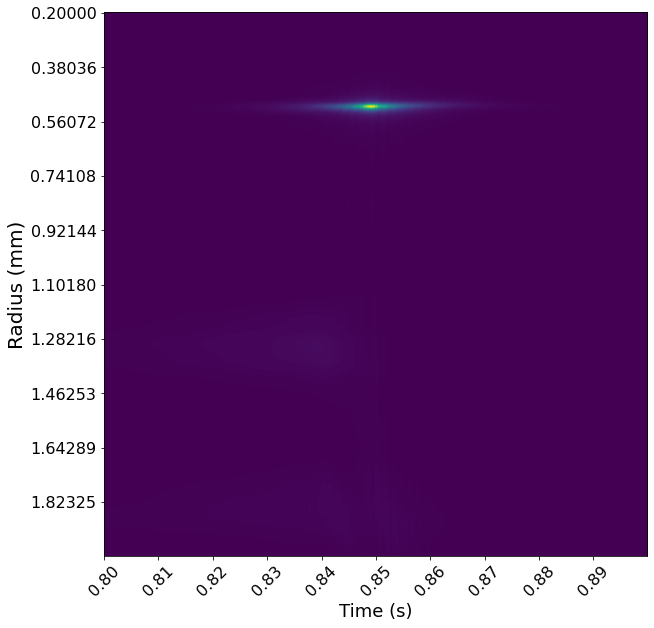}{.4\textwidth}{(c) $6$ $mm$ nozzle}}
%
 \caption{Bubblegrams for a single bubble with multiple nozzle diameters}
 \label{fig:fig8}
\end{figure}

The bubblegrams obtained from the preprocessed signals appear in figure \ref{fig:fig8}.

The Maximum Posterior (MAP) estimates for the bubbles radii were $0.6509$, $0.4273$ and $0.5102$ $mm$ for the $2.5$, $4$ and $6$ $mm$ nozzles respectively. By comparing this estimates to the pictures in figures \ref{fig:fig8} we find that the estimates are accurate, considering the deformation of the bubbles in each snapshot.


\subsubsection{Bubble plumes}

To obtain the acoustic emission of bubble plumes at higher flow rates the diving tank was used. The nozzle of the pneumatic system was positioned at the bottom of the tank, and several runs were conducted with different nozzle diameters and gas flow rates. For each run, we open the flow controller and let the gas escape for two minutes. Then we close the controller and wait for one minute to open it again.

In this section we are interested in studying both the spectrogram and bubblegrams for different values of the nozzle diameter and gas flow rates. We analyze signals obtained with diameter $D \in \{2.5, 6, 11\}$ $mm$ and flows of $F \in \{2, 5, 10\}$ $l\min$, ie, we analyze $9$ different sections of the experimental signal. 

\begin{figure}[H]
\figline{\fig{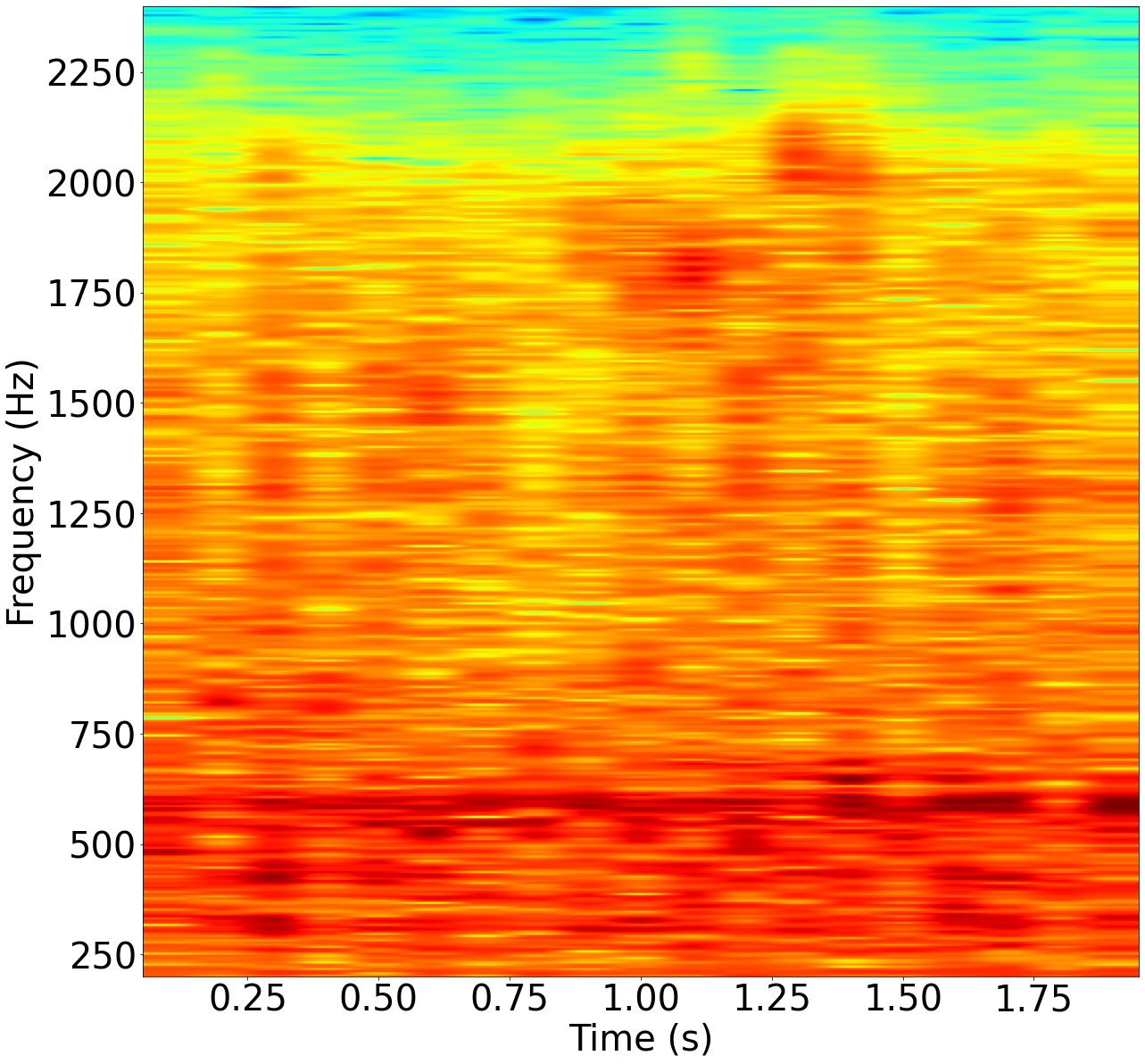}{4cm}{(a) $d = 2.5$ $mm$; $f = 2$ $l/min$}
\fig{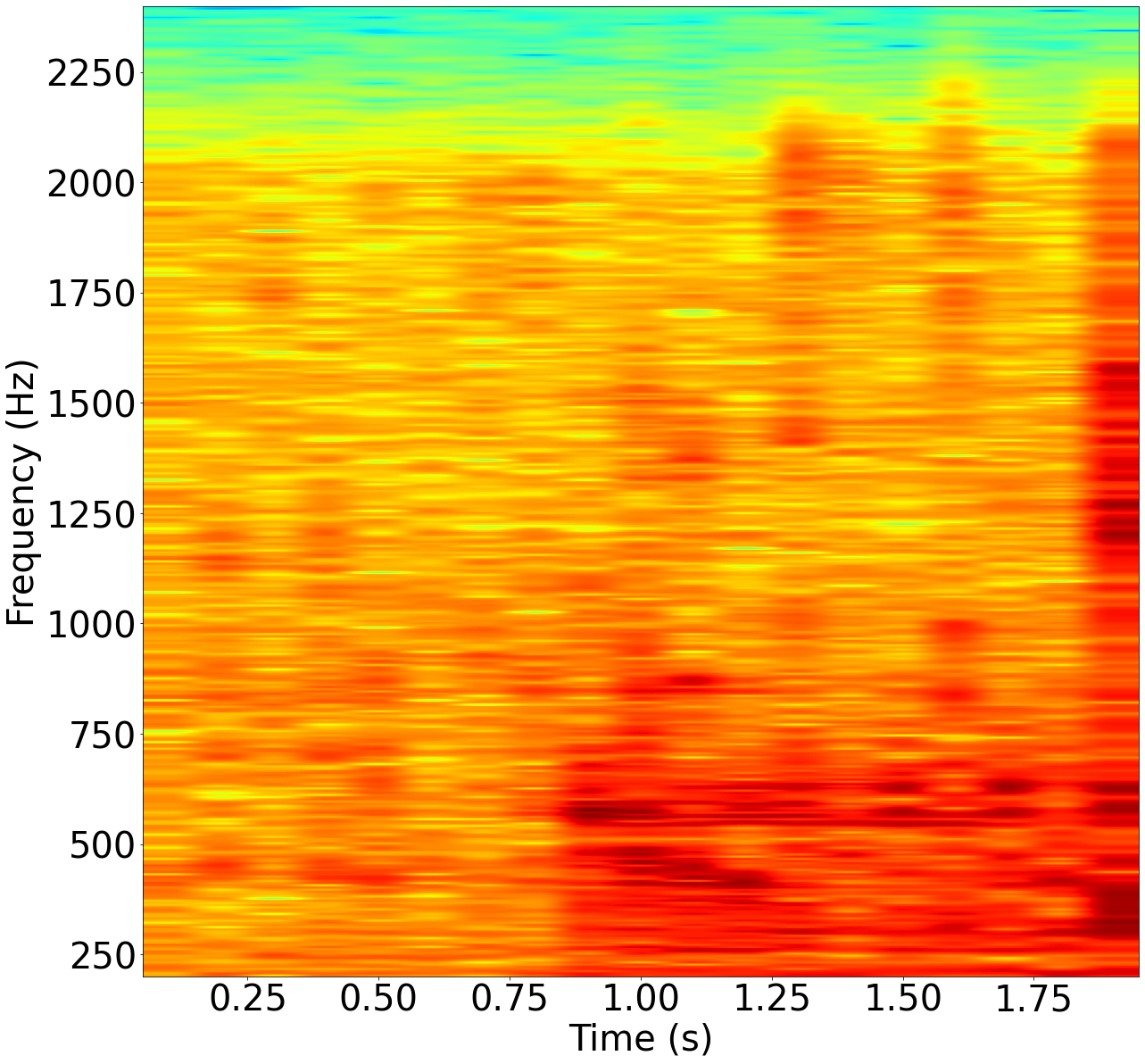}{4cm}{(b) $d = 6$ $mm$; $f = 2$ $l/min$}
\fig{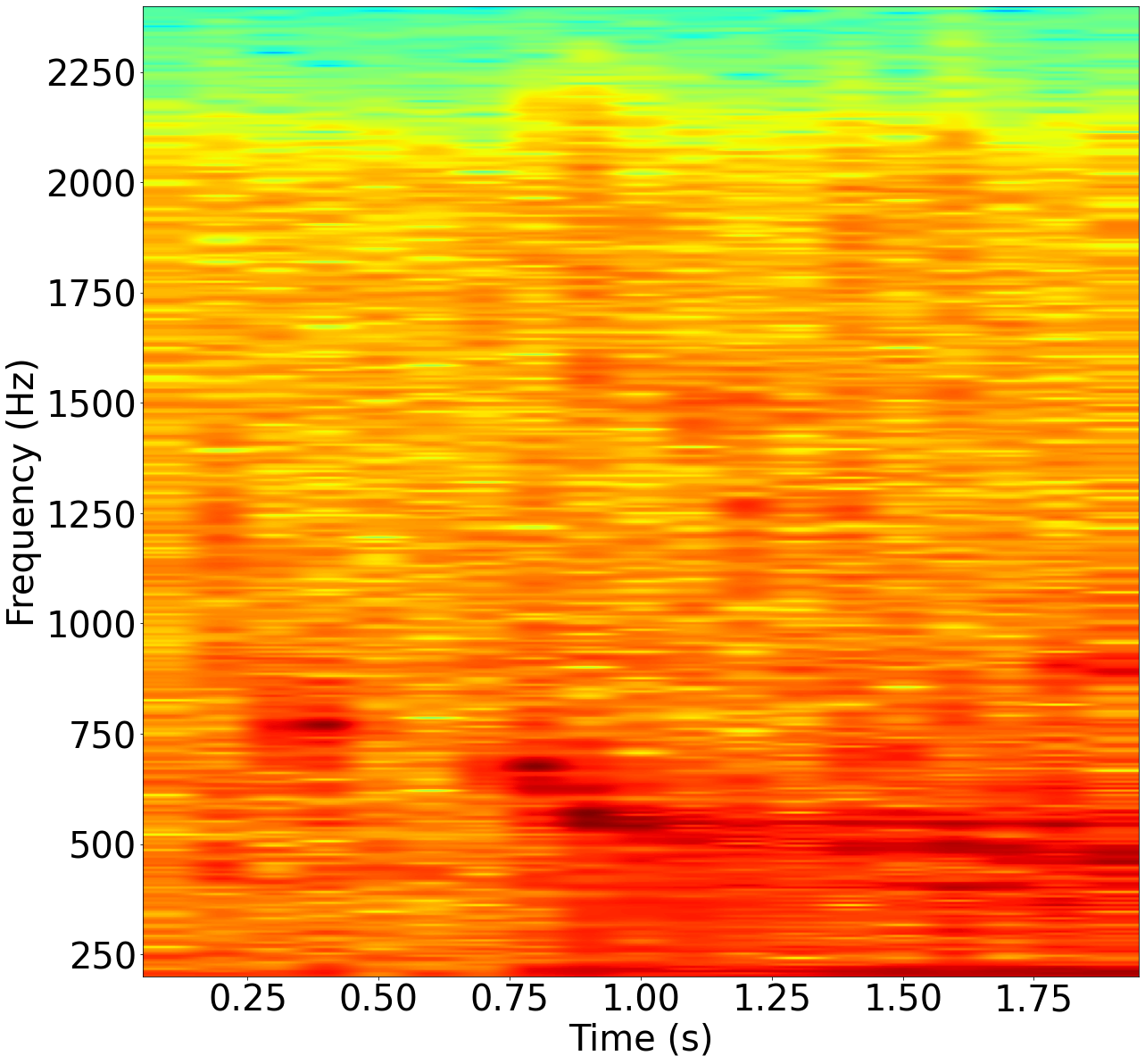}{4cm}{(c) $d = 11$ $mm$; $f = 2$ $l/min$}} 
\figline{\fig{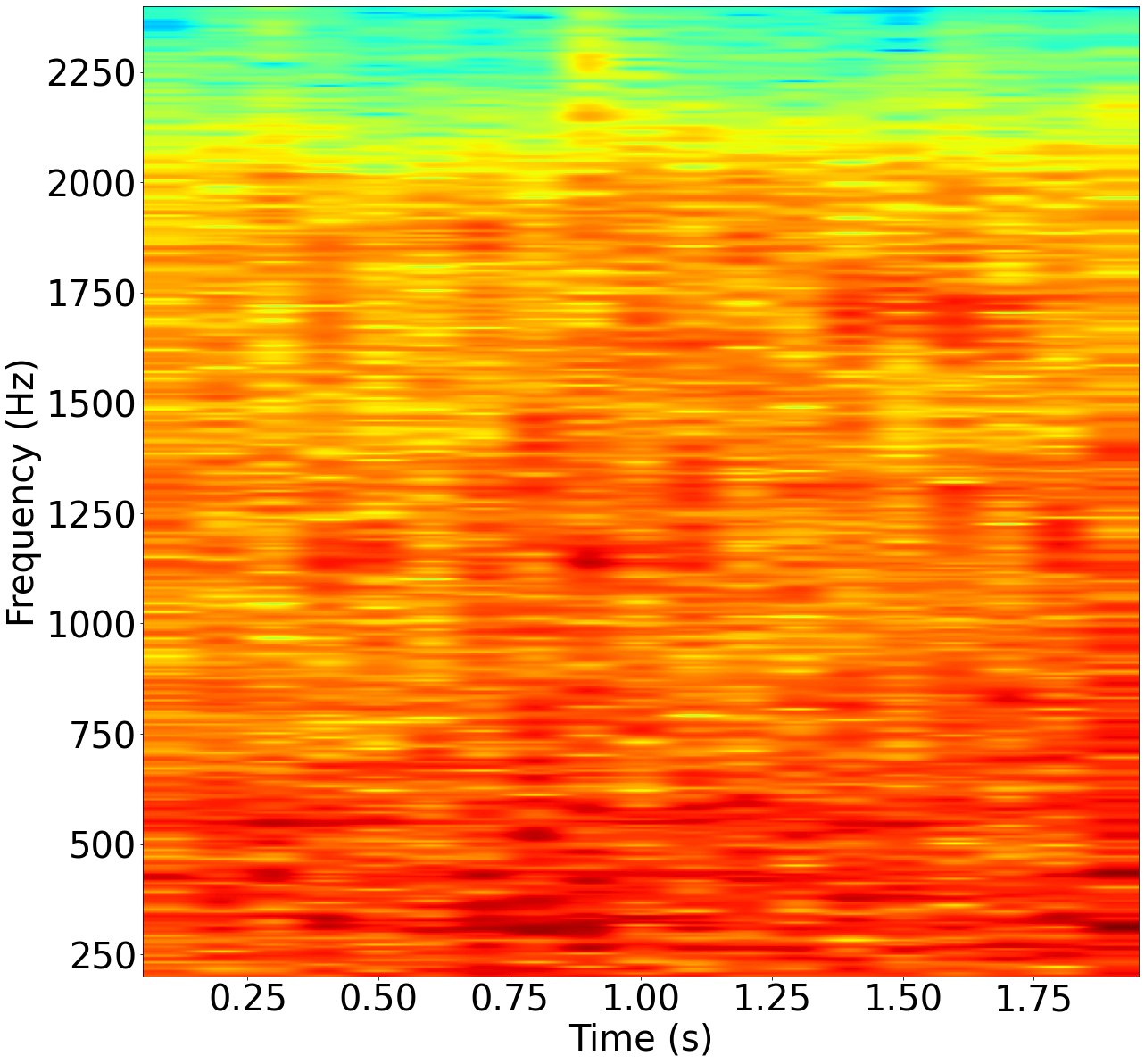}{4cm}{$d = 2.5$ $mm$; $f = 5$ $l/min$}
\fig{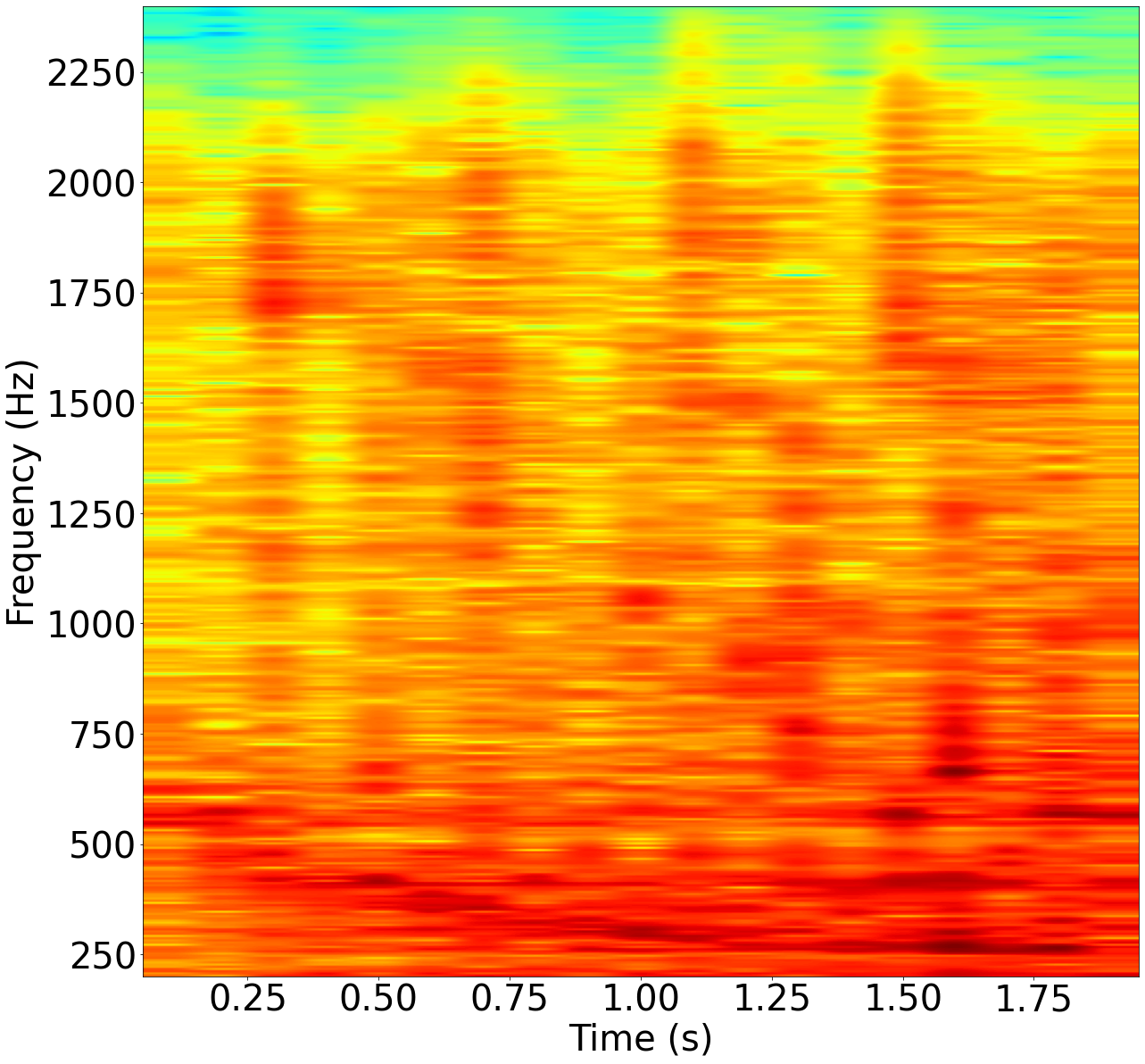}{4cm}{$d = 6$ $mm$; $f = 5$ $l/min$}
\fig{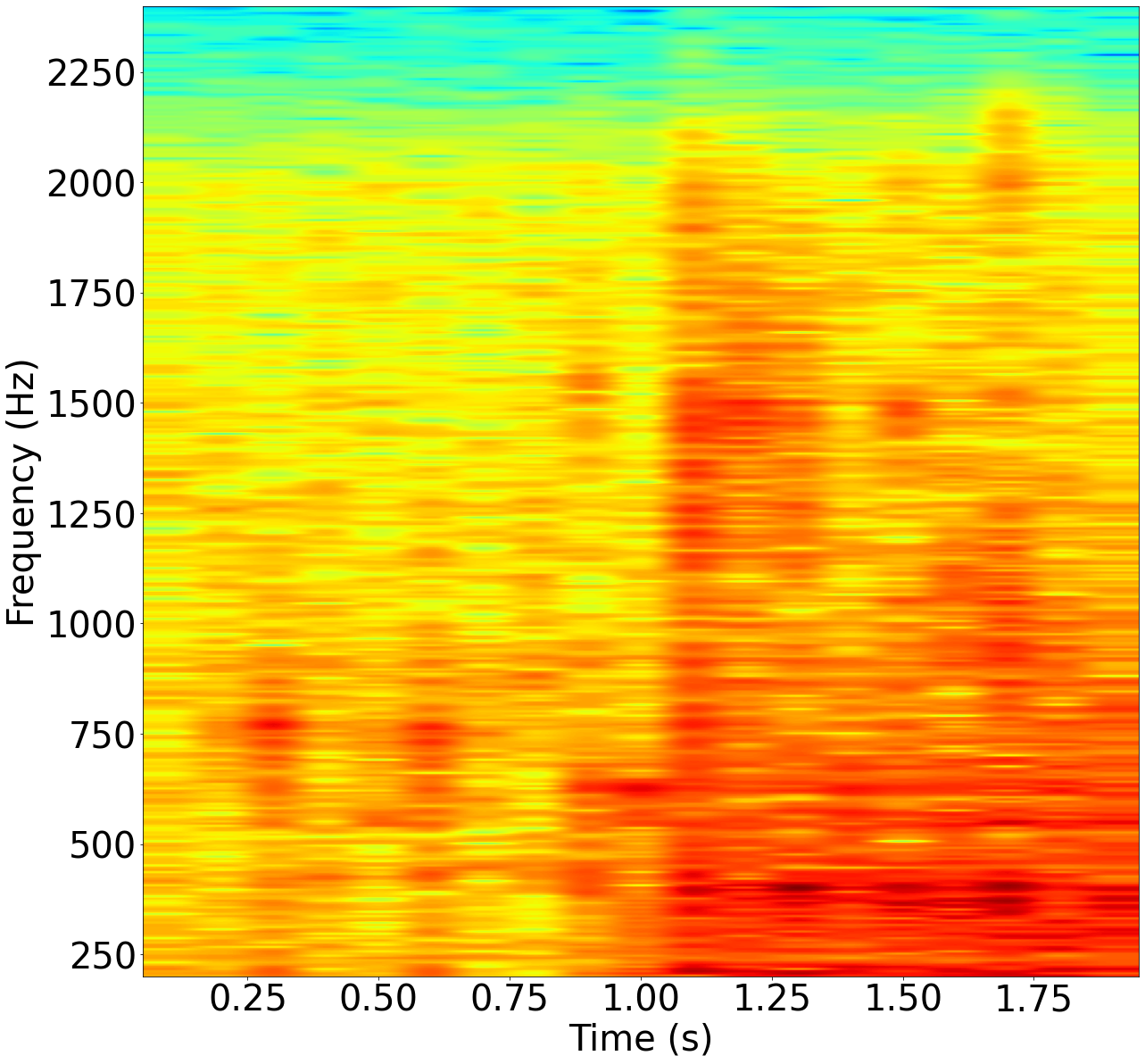}{4cm}{$d = 11$ $mm$; $f = 5$ $l/min$}}
\figline{\fig{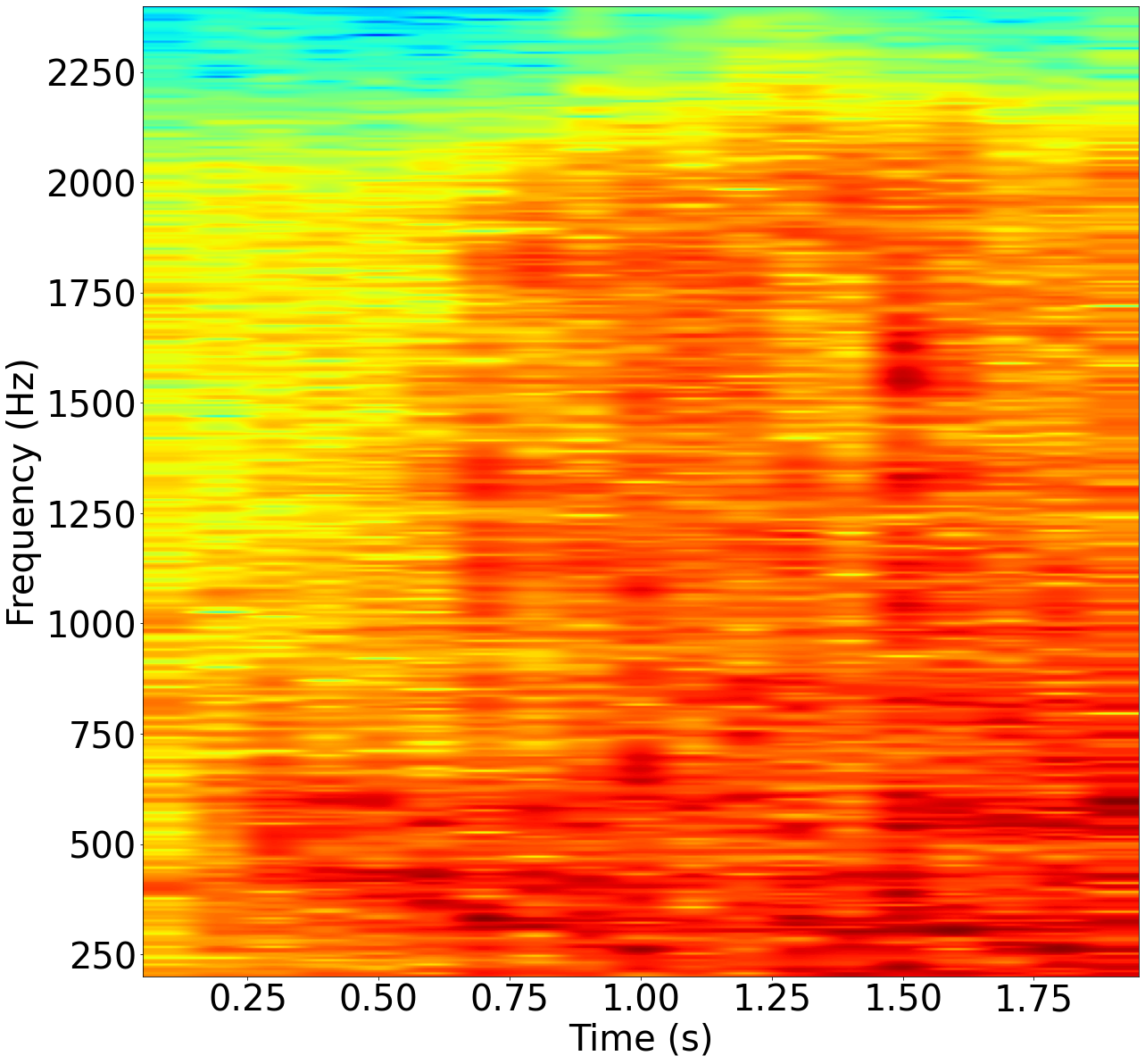}{4cm}{$d = 2.5$ $mm$; $f = 10$ $l/min$}
\fig{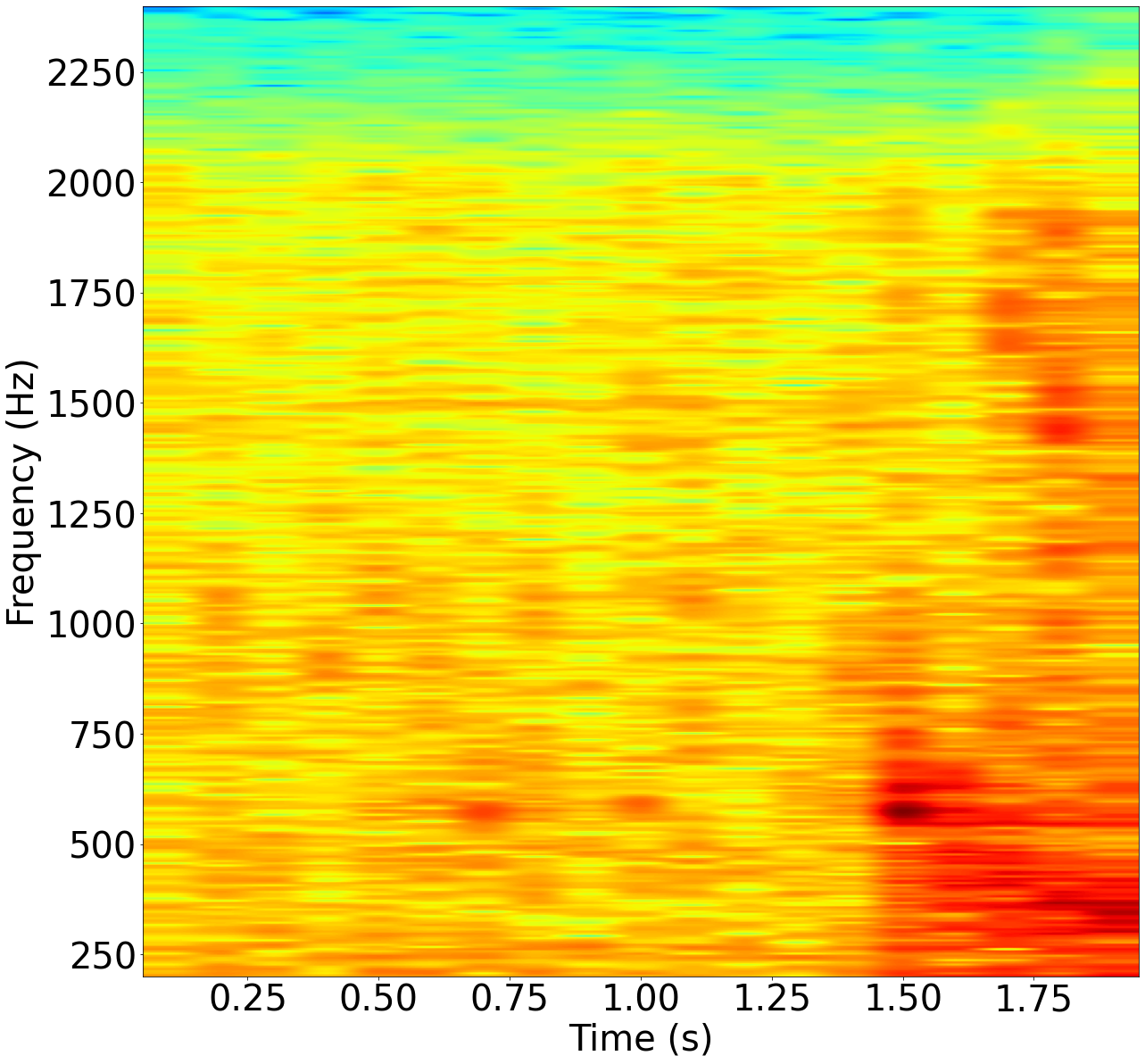}{4cm}{$d = 6$ $mm$; $f = 10$ $l/min$}
\fig{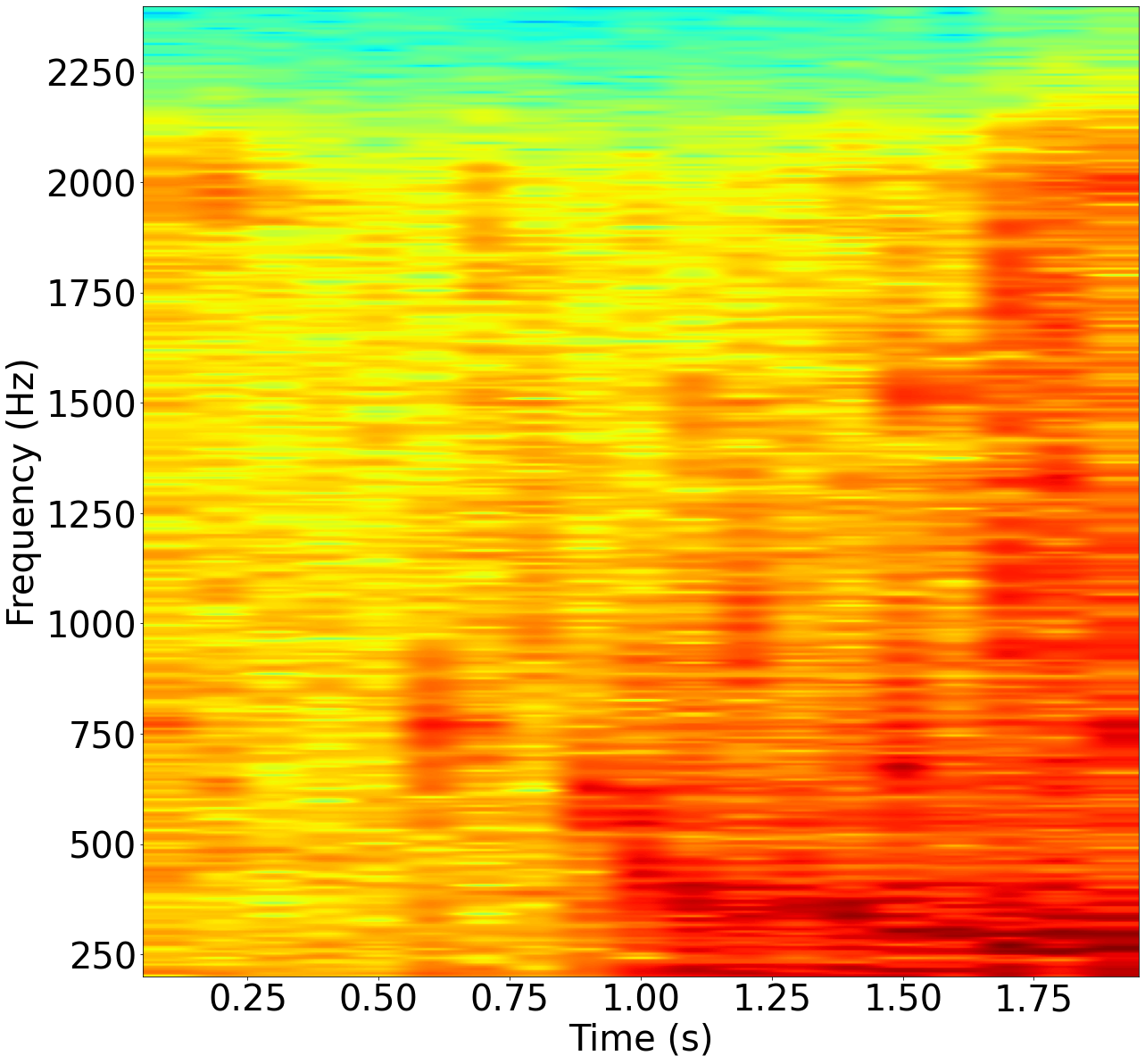}{4cm}{$d = 11$ $mm$; $f = 10$ $l/min$}}
 \caption{Full spectrograms for bubble plume data.}
 \label{fig:fig9}
\end{figure}

To conduct our analysis we extract small sections ($2$ seconds) of the signals and start by obtaining the spectrograms of each section. These spectrograms appear in figure \ref{fig:fig9}.

The spectrograms show that the signal's energy is concentrated around smaller frequencies (between $250$ and $1000$ $Hz$). However, for the reasons already exposed, the graphs are noisy and do not allow easy or direct estimation of the probable sizes of the bubbles.

\begin{figure}[!ht]
\figline{\fig{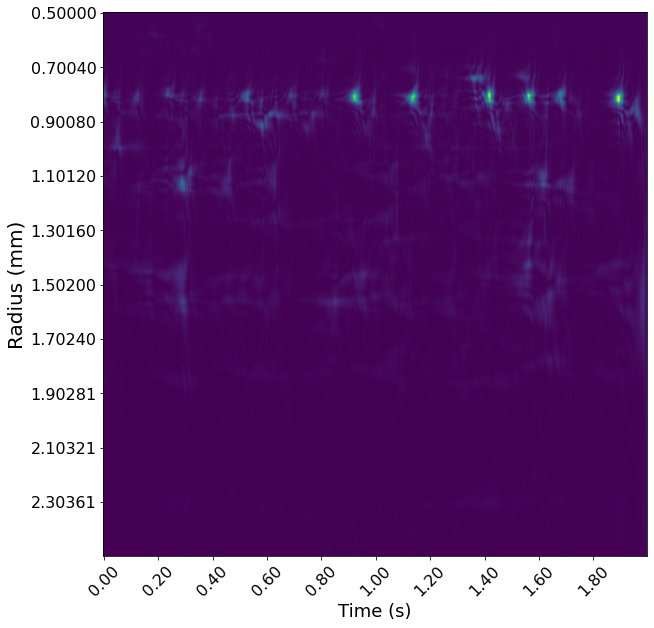}{4cm}{(a) $d = 2.5$ $mm$; $f = 2$ $l/min$}
\fig{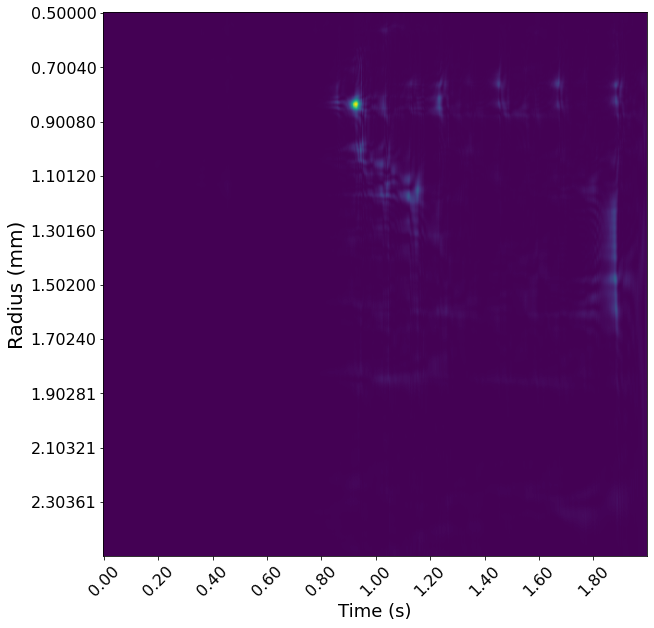}{4cm}{(b) $d = 6$ $mm$; $f = 2$ $l/min$}
\fig{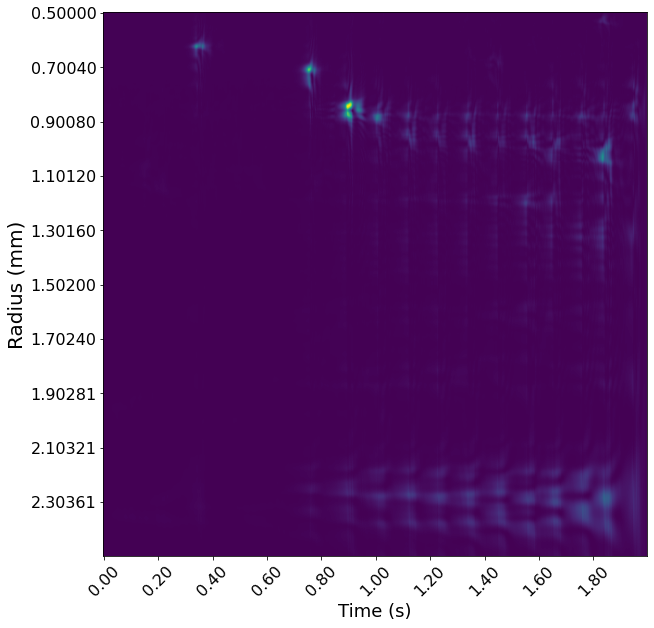}{4cm}{(c) $d = 11$ $mm$; $f = 2$ $l/min$}} 
\figline{\fig{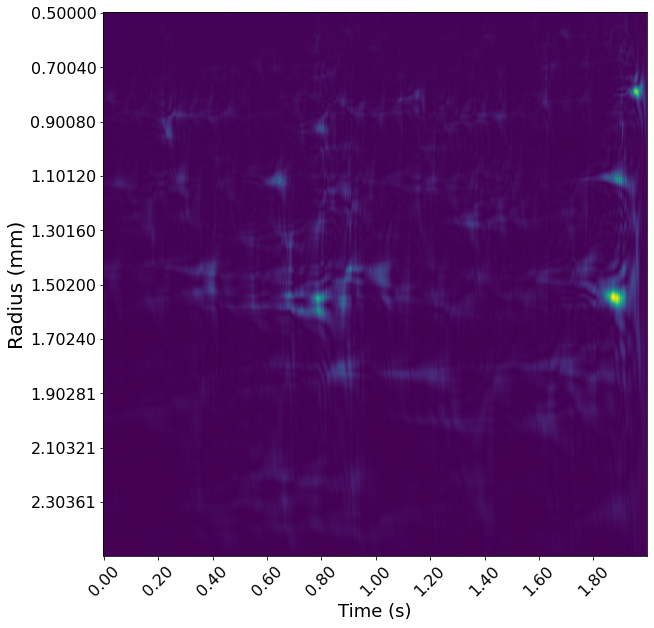}{4cm}{$d = 2.5$ $mm$; $f = 5$ $l/min$}
\fig{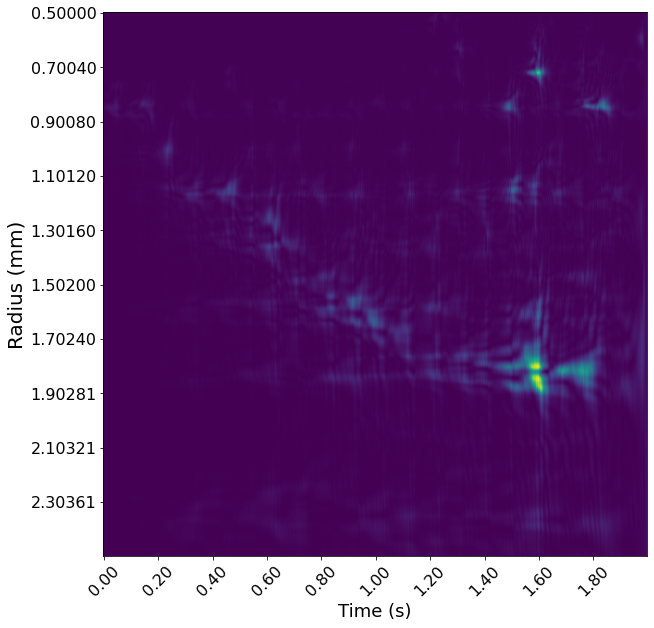}{4cm}{$d = 6$ $mm$; $f = 5$ $l/min$}
\fig{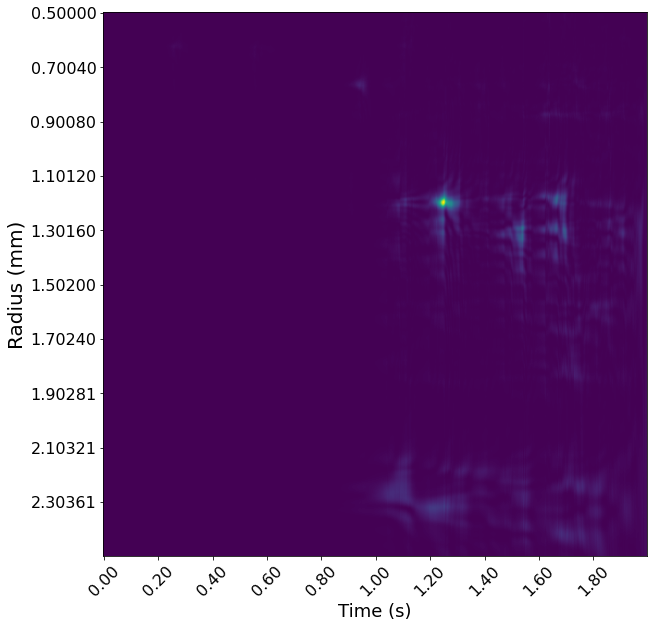}{4cm}{$d = 11$ $mm$; $f = 5$ $l/min$}}
\figline{\fig{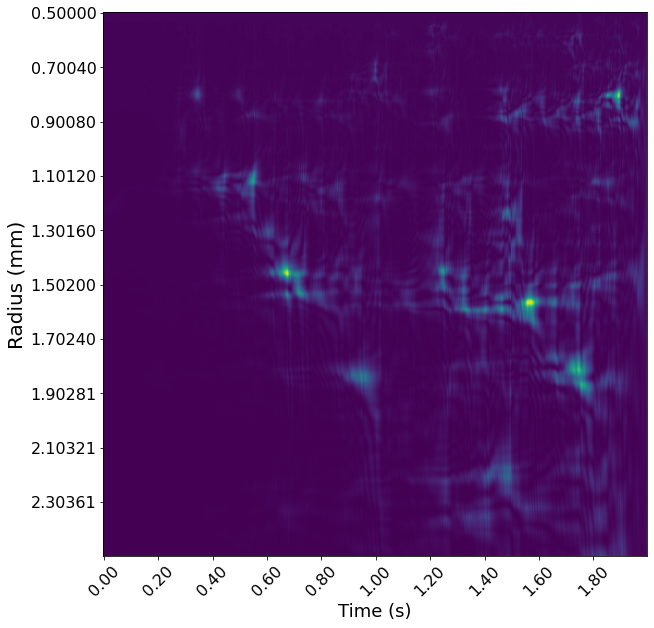}{4cm}{$d = 2.5$ $mm$; $f = 10$ $l/min$}
\fig{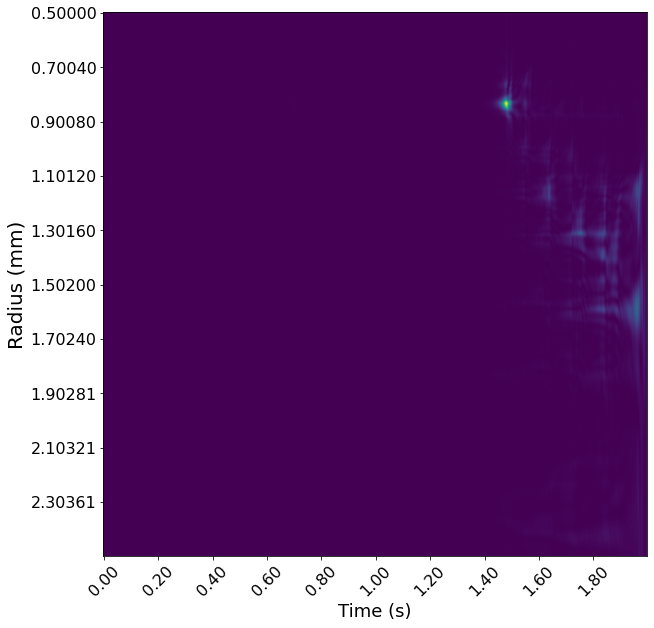}{4cm}{$d = 6$ $mm$; $f = 10$ $l/min$}
\fig{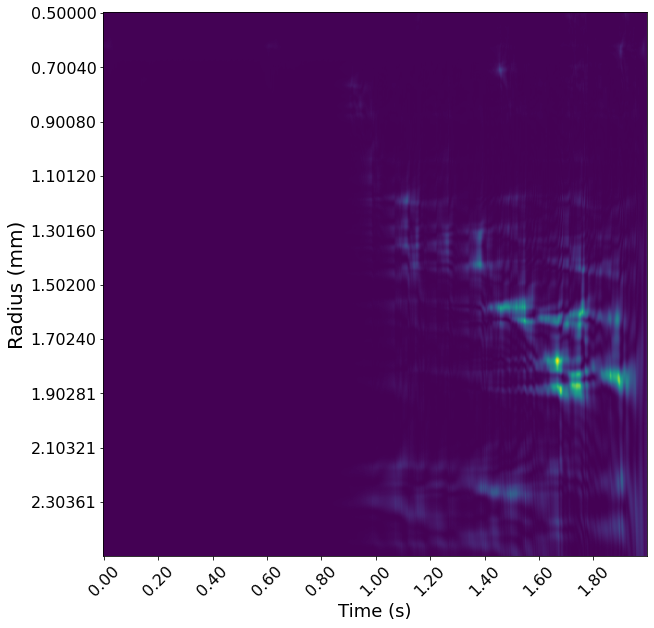}{4cm}{$d = 11$ $mm$; $f = 10$ $l/min$}}
  \caption{Bubblegrams for bubble plume data.}
  \label{fig:fig10}
\end{figure}

The bubblegrams of figure \ref{fig:fig10}, on the other hand, are much clearer and easier to inspect. As we can see, higher flow rates are associated with bigger bubbles.

The bubblegram is a useful representation of the signal, directly obtaining a probability distribution over bubble sizes. From the figures we can observe that higher gas flow rates create bubble plumes with more diverse sized bubbles (we see that from the regions with relevant evidence, colored blue, but far from the more concentrated regions that show the typical bubble size in each case). Also the time between the formation of each bubble changes according to the gas flow.

The gas flow appears to have a more evident effect on the bubble size distribution than the nozzle diameter, as we can see by comparing figures in the same row (i.e., signals obtained with the same flow) and then comparing figures in the same column (signals obtained with the same nozzle diameter). However both gas flow and nozzle diameter are positively correlated with bubble sizes, even though the flow rate has a stronger effect.

\begin{figure}[H]
\figline{\fig{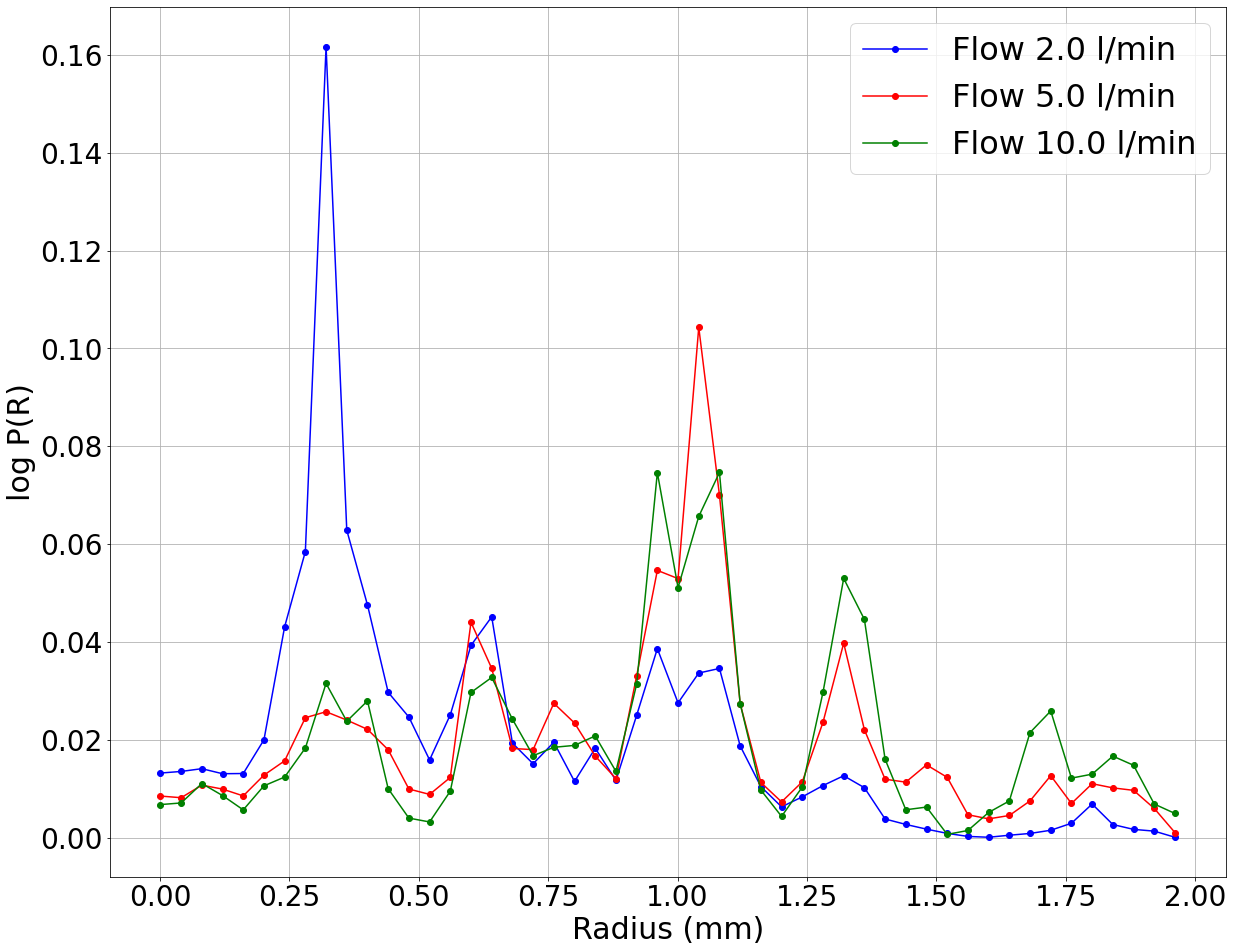}{.4\textwidth}{$d = 2.5$ $mm$}
\fig{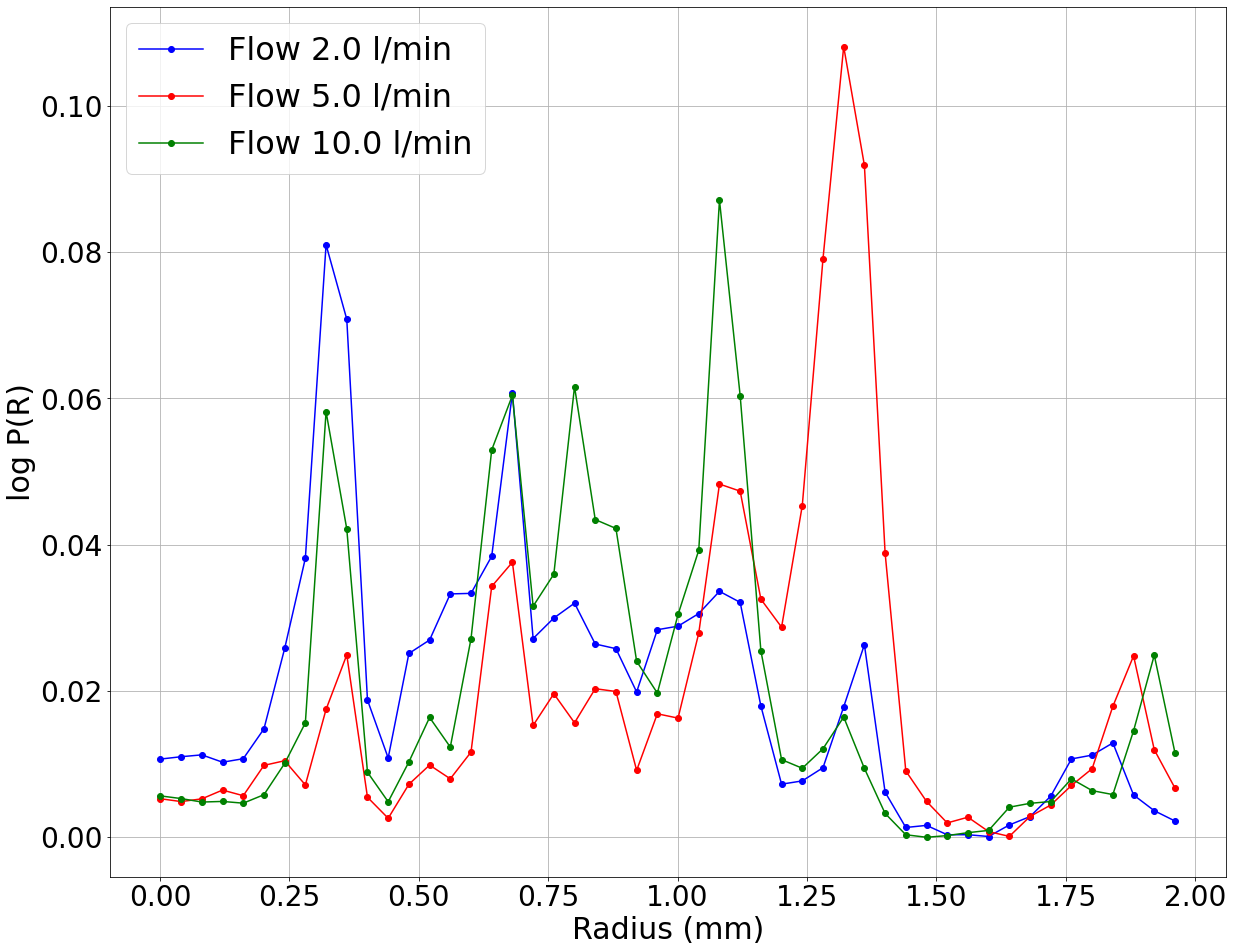}{.4\textwidth}{$d = 6$ $mm$}}
\figline{\fig{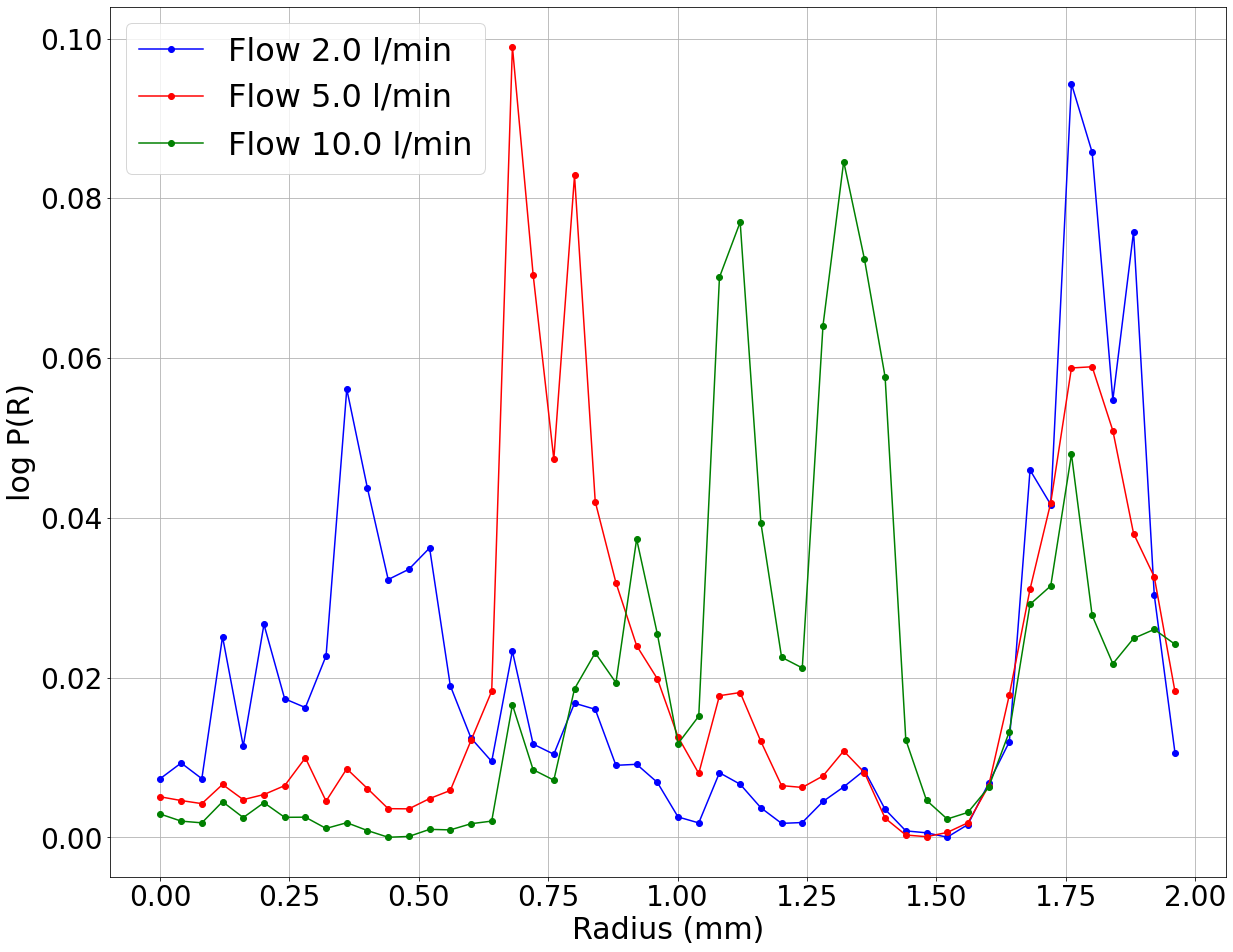}{.4\textwidth}{$d = 11$ $mm$}}
 \caption{Log-posterior density for bubble sizes, fixed flow}
 \label{fig:fig11}
\end{figure}

By aggregating and normalizing the log-posterior for the bubble sizes we arrive at the approximate density for the distribution of bubble sizes in the entire signal. In figure \ref{fig:fig11} we plot these distributions for a fixed flow rate, and varying nozzle diameter.

With these figures we are able to observe more directly the effect of both nozzle diameter and flow rate in the estimated bubble size distribution. 

Our results indicate that this relationship is complex. For instance, from figure \ref{fig:fig11} we observe first of all that the smallest nozzle induces more homogeneous bubbles populations, since the curves are close to being unimodal (specially for the lowest flow, blue line). For this smallest flow, the mode radius is also the smallest, around $0.25$ $mm$; the two bigger flows have very similar curves.

Increasing the nozzle diameter we observe that the bubble populations start to break down into subpopulations with different sizes (that is, the curves start to show large deviations from unimodality). The curve for the smallest flow (blue lines in all three plots) has always a prominent peak around $0.25$ $mm$, but as the nozzle size increases bigger bubbles can also be found according to our analysis. The intermediate and higher flows that we tested behave less monotonically; for a flow of $5$ $l/min$, when the diameter is set at $6$ $mm$ (figure \ref{fig:fig11}, red line) there appears a high peak around a radius of $1.25$ $mm$, indicating that increasing the nozzle allows the formation of bigger bubbles. However, when the diameter increases to $11$ $mm$, a subpopulation of bubbles with radii around $0.75$ $mm$ appears, along with another subpopulation of bubbles with raddi around $1.75$ $mm$. Therefore, for the bubble plumes with intermediate flow, the bubbles increase in size with nozzle diameter, but for the biggest diameter a population of small bubbles is also formed.

The curves obtained from the analysis of the samples with biggest flow (green line in all three subplots) all have an important peak around $1$ $mm$. Increasing the nozzle diameter from $2.5$ to $6$ $mm$ causes the appearance of subpopulations of smaller bubbles (peaks around $0.25$ and $0.7$ $mm$, green line in figure \ref{fig:fig11}). When the nozzle increases to $11$ $mm$ (figure \ref{fig:fig11}) the peak around smaller bubbles is much attenuated, and is replaced by peaks around $1.25$ and $1.75$ $mm$.


\section{Conclusion}\label{sec:conc}

We applied a known methodology of probabilistic (Bayesian) signal processing to the analysis of the acoustic emission of air bubbles in water. Our model directly expresses the acoustic signal in the time domain as a function of the bubble radii and time of occurrence, thus providing a useful technique to analyze acoustic signals containing the emission of bubble plumes.

Testing our model on laboratory generated data we found evidence that the model is accurate, providing good estimates of the size of a single bubble from its acoustic emission.

To analyze the acoustic emission of bubble plumes we propose to evaluate the posterior of the single bubble model in small temporal segments of the signal in the time domain. This is done in a similar way as the usual Short Time Fourier Transform, and can be interpreted as imposing an informative prior over the time of occurrence of a single bubble. This evaluation, performed over a grid of values for both the time of occurrence and bubble radius, generates a visual representation we called \emph{bubblegram}, that is also similar to the usual spectrogram, but is built on a full probabilistic setup that embodies a physical model relating the bubble size to the parameters of an exponentially decaying sinusoidal. As such, the bubblegram provides a representation of the signal that is more accurate than the spectrogram, and can be helpful in the study of the bubble sizes in bubble plumes by passive acoustic methods.

\bibliography{bibliografia}

\end{document}